\documentstyle[preprint,aps]{revtex}
\begin{document}
\draft
\title { 
 Numerical study of the two dimensional Heisenberg model  by  \\
 Green Function Monte Carlo at fixed number of walkers} 
\author{Matteo Calandra Buonaura and Sandro Sorella} 
\address{
Istituto Nazionale di Fisica della Materia and International
School for Advanced Study, Via Beirut 4, 34013 Trieste, Italy}
\date{\today}
\maketitle
\begin{abstract}
We describe in detail a simple and     
 efficient Green Function Monte Carlo technique 
for computing both the ground state energy 
 and the ground state properties by the ``forward walking'' scheme.
The simplicity of our reconfiguration  process, used to 
maintain the walker population constant,  
 allows to control any source of systematic error  in a  rigorous and 
systematic way.  
We apply this method  to the Heisenberg 
model and obtain accurate and reliable 
estimates  of the ground state 
energy, the order parameter and the  static spin structure factor  $S(q)$ 
for several momenta. For the latter quantity we also find very good agreement
with available experimental data on the  $La_2CuO_4$ antiferromagnet. 
\end{abstract}
\pacs{75.10.Jm,75.40.Mg,75.30.Ds}
\widetext
\section{Introduction}
After  almost one decade from the discovery of High-Tc 
superconductivity we have certainly  understood much more  about 
magnetism rather than superconductivity. 
In particular since almost 
all the stechiometric  compounds  of High-Tc superconductors   are 
good antiferromagnets, 
  well described by the two dimensional  Heisenberg model (HM),
  from the very beginning a  strong numerical effort has been  devoted  to 
the simulation of this model.\cite{reger,ceperley} 
The  HM is  defined by the following Hamiltonian:
\begin{equation} \label{heisenberg}
H_J= J \sum_{<i,j>} \vec S_i \vec S_j 
\end{equation}
where the spin one half vectors $\vec S_i$   satisfy $\vec S^2 =3/4$ 
and  $J$ is the nearest neighbor 
antiferromagnetic superexchange coupling, connecting nearest neighbor  pairs 
$<i,j>$. Henceforth  periodic boundary conditions are assumed in a 
finite square lattice with $N_a= l \times l$ sites.

Although a rigorous proof that this model has long range antiferromagnetic 
order
in two spatial dimension is still lacking, there is a general consensus 
that long range order exists even in this interesting case. 
In other words its  properties should be very  well understood by
the simple spin-wave theory, which assumes long range antiferromagnetic 
order in the ground state.\cite{anderson}  

In this  work we give accurate  ground state properties of the 
HM using a new and more  efficient version 
of the Green Function Monte  Carlo (GFMC) technique, first applied 
on a lattice by Trivedi and Ceperley, some years ago\cite{ceperley}. 

 With the present  scheme we also estimate the ground state energy of the
HM  on the square lattice to be $-0.669442 J \pm 0.000026 J $
 slightly different, but more accurate  of the previous GFMC estimates. 
Analogously we obtain for the   antiferromagnetic order parameter  
the value $m=0.3077 \pm 0.0004$, 
consistent with other numerical estimates    but with a very accurate 
control of the finite size effects.
We discuss also our results for the static spin structure factor 
in view of the  recently proposed theory for the  
 finite size scaling in a quantum antiferromagnet.\cite{ziman}
In particular we verify  that,  as a remarkable prediction of 
 the theory,  the small $q$ 
behavior of this function behaves as $ S(q) = \chi c |q|$, where $\chi$ 
is the magnetic spin susceptibility and $c$ the spin wave velocity of the HM.
This relation is particularly important as this function is experimentally 
detectable in neutron scattering experiments.\cite{allen}

Let us discuss  now  the technical part of our work,
 which is based on the GFMC, as we have mentioned before.
As is well known this technique   allows to
sample statistically 
 the ground state of a many body Hamiltonian $H$ by a set of
walkers $(w_i,x_i)$  which represent vectors  $ w\,x$
of a large (or even infinite)  Hilbert space.  
The set of all 
configurations $x$ spans  a normalized and complete basis. 
The aim of this approach is to 
 sample statistically the  ground state of $H$,  by a large population of
walkers.\cite{linden}  As it will be described later on, 
in the finite dimensional case,  the GFMC  on a lattice   
is based on a statistical application  of the Hamiltonian matrix-vector product
$ w^\prime_i  x_i^\prime  \to (-H) w_i x_i$ to the walker configurations
 $\left\{w  x\right\} _i $  , thus  filtering out, after many iterations the
desired population distribution for the ground state.    
In this statistical iteration  however, the walker weights $w_i$  
increase or decrease exponentially so that after a few iterations most of the 
walkers have negligible weights 
and some kind of reconfiguration
becomes necessary to avoid large statistical errors. 
The process to eliminate the irrelevant walkers or generate copies of 
 the important
ones   is called ``branching''. 
This scheme is in principle exact 
 only if the population of walkers is let increase or decrease without any 
 limitation. Any reconfiguration of the population size may in fact introduce 
some spurious correlation between the walkers that may affect the 
statistical sampling of the ground state.  
 In practice for long simulation it is always
necessary to control the population size, as,  otherwise, one easily 
exceeds  the maximum allowed 
computer memory. 
 This control of the walker population size 
 may  introduce some kind of bias
that  vanishes quite slowly for
infinite number of walkers.\cite{linden,kalos}  
In this case  only by  performing  several runs with different numbers of 
walkers one may in principle estimate the size of the residual bias. 

Following the Hetherington's work\cite{hether},
 we  define here an efficient  reconfiguration process  
at fixed number $M$ of walkers,  with a rigorous control 
of the bias and without need of the conventional branching scheme.  

In the last sections  we present the results obtained for  
 the HM up to $N_a=16\times 16$ lattice size,
 together with some numerical test on a small $N_a=4 \times 4$ 
lattice where an  accurate  
numerical solution is available  by exact diagonalization. 
Previous calculations on this model, using the Green Function Monte Carlo
technique, were performed either without correcting the bias\cite{ceperley}
and controlling it for small lattices with a large  number of walkers, or 
by correcting the bias  in a way, which is probably correct, but it is not
possible to prove rigorously.\cite{runge}

\section{The GFMC technique}
 
In the following sections    we describe in detail  how 
 to evaluate  the maximum eigenvector of 
a   matrix $H_{\displaystyle x^\prime,x}$ with all positive definite 
 matrix elements, using a
stochastic approach. 
Clearly in any physical problem,  described by an Hamiltonian $H$,  the most
interesting eigenvalue is the lowest one: the ground state energy. 
This is however  just a matter of notations,  
as the ground state of $H$ represents the maximum excited state of $-H$. 
 In the following, for simplicity, 
 we assume  to change  the sign of $H$ so that 
 the physical ground state is, in this notations, 
 the maximum eigenvector of $H$.
More important is instead the restriction of  positive definite matrix
elements $H_{\displaystyle x^\prime, x}$, which drastically constrains  
the  class of  Hamiltonians that can be treated   with this method,  
without facing the old but still unsolved  ``sign problem''.
Whenever the  Hamiltonian has matrix elements with arbitrary sign
schemes like ``Fixed nodes approximation'', and their recent developments to
finite lattices, are possible within  
the GFMC  method.\cite{bemmel,ceperley1} 
Of course if negative signs occur only in the diagonal elements of $H$ a simple
change  of the hamiltonian $H \to H_{x^\prime,x}  + \lambda \delta_{\displaystyle
x^\prime,x} $, will not change the ground state but the Hamiltonian will satisfy
the condition $H_{x^\prime,x} \ge 0$  for  large enough shift $\lambda$.
For instance the Heisenberg hamiltonian can be easily casted in the
previous form   as it had been previously shown.\cite{ceperley}

From a general point of view 
  the ground state of $H$ can be obtained  by applying the
well known power method: 
\begin{equation} \label{power}
|\psi_{\displaystyle 0}> = \lim_{\displaystyle L\to \infty} H^L |\psi_T>
\end{equation}
where the equality holds up to  (infinite) normalization, and $|\psi_T>$ is a
trial state non orthogonal to the ground  state $|\psi_{\displaystyle 0}>$.

In the following  a simple stochastic approach is described 
  for evaluating 
  the state $H^M|\psi_T>$.
To this purpose we define a basic element of this stochastic approach : the 
so called walker. A walker is determined by an index $x$ corresponding to a
given element $|x>$ of the chosen basis and a weight $w$. 
The walker 
``walks'' in the  Hilbert space of the matrix $H$ and assumes 
a configuration $w x$ according to a given probability distribution $P(w,x)$.

The task of the Green function Monte Carlo approach is to define a Markov  
process, yielding after a large number $n$ of  iterations a  probability
distribution $P_n (w,x)$  for the walker which determines the ground state
wavefunction $\psi_{\displaystyle 0}$. To be specific in the most simple
formulation one has:
 $$\int dw \, w\, P(w,x)= <x|\psi_{\displaystyle 0}>$$.

\section{Single walker formulation}
In the following 
the distribution $P(w,x)$ is sampled by a finite number $M$ of walkers.
Let us first consider the simpler case $M=1$.
In order to define a statistical implementation of the matrix multiplication 
$|x> \to  H |x>$, the standard approach is first  to determine the
hamiltonian matrix elements $H_{\displaystyle i,x}$ connected to $x$ which are 
different from zero. Then a new 
 index $x^\prime$ is chosen for the  walker among 
the indices $i$  according to the probability determined by 
\begin{equation} \label{markov}
p_{\displaystyle i,x}= H_{\displaystyle i,x}/b_{\displaystyle x}
\end{equation}
 where $b_{\displaystyle x}=\sum_i H_{\displaystyle i,x}$ 
has been introduced in order to satisfy the normalization  condition
$\sum_i p_{\displaystyle \displaystyle i,x}=1$.
This simple iteration scheme to go from a configuration $x$ to a new
configuration $x^\prime$  is easily implemented but is not 
sufficient to determine stochastically the matrix-vector product $H x$.
The full matrix is a product of a stochastic matrix $p_{\displaystyle
i,x}$ and a diagonal one $b_{\displaystyle x}$:
\begin{equation} \label{relpx}
H_{\displaystyle i,x}= p_{\displaystyle i,x} b_{\displaystyle x} 
\end{equation}
As it is intuitively clear the diagonal matrix $b_{\displaystyle x}$, 
 not included in the stochastic process,  is very easily
determined  by a scaling of  the weight $w$ of the walker:
\begin{equation} \label{scalw}
w^\prime \to b_{\displaystyle x} w
\end{equation}
The two previous updates, the stochastic one (\ref{markov}) and the
deterministic one (\ref{scalw}) 
 define a new walker $w^\prime, x^\prime$ in place of the
``old'' walker $w,x$, i.e. they determine a Markov process. 

At this point it is important  to understand   the evolution of the probability
distribution $P(w,x)$  after such process.
 A subscript $n$ to this function $P$ will indicate   the number 
of iterations of the Markov process.
 The probability evolution $P_n \to P_{\displaystyle n+1}$ 
is easily determined by 
\begin{equation}  \label{procp}
P_{\displaystyle n+1}(w^\prime,x^\prime)= \sum_x p_{\displaystyle x^\prime,x} P_n( w^\prime/b_{\displaystyle x},x) /b_{\displaystyle x}
\end{equation}  

The equation (\ref{procp}) allows to determine, 
 by simple iteration, what is the
probability to find a walker in a given configuration $w,x$ after many steps.
However the  evolution   $P_n$ from the initial distribution $P_0$ is more clear 
and transparent in terms of its momenta  over the weight
variable $w$: 
\begin{equation} \label{defmomk}
G_{\displaystyle k,n} (x)= \int dw \, w^k \, P_n(w,x) 
\end{equation}
In fact it is straightforward to verify, using  Eq.(\ref{procp}) , that :
\begin{equation} \label{defitg} 
G_{\displaystyle k,n+1} (x^\prime) = \sum_{\displaystyle x}  p_{\displaystyle
x^\prime,x}
 b_{\displaystyle x}^{\displaystyle k} G_{\displaystyle k,n} (x)
\end{equation}

In particular for $k=1$  the first momentum of $P$ determines  the
full  quantum mechanical information, as $G_{\displaystyle 1,n} (x^\prime)  = 
( H^{\displaystyle n} ) _{\displaystyle x^\prime,x} G_{\displaystyle 1,0} (x) $,
implying that $G_{\displaystyle 1,n}(x)$, by Eq.(\ref{power}), 
converges to the ground state   of the hamiltonian $H$.

 By iterating several  times even a single walker , the resulting configuration 
$w,x$ will be distributed according to the ground state of $H$ and by
sampling a large number of independent configurations we can evaluate for
instance the ground state energy:
\begin{equation} \label{energy}
E_{\displaystyle 0}=  { < w b_{\displaystyle x} > \over < w > }
\end{equation}
where the brackets $< >$ indicate the usual stochastic average,
 namely averaging
over  the independent configurations.

This in principle  concludes the GFMC scheme.
However
  the weight $w$ of the walker grows exponentially with $n$ (simply as a 
result of $n$ independent products) and can
assume very large   values,  implying diverging variance in the above averages.

In the next sections we describe in detail this problem and a way to solve it 
with a {\em fixed} number of walkers. 
\section{ Statistical average during the Markov process}
The configuration $x_n$ that  are generated in the Markov process are 
distributed  after long time according 
to the  maximum right eigenstate $R(x)$ 
of the matrix $p_{\displaystyle x^\prime ,x }$
 ( simply because  
$G_{\displaystyle n,0} (x) = \sum_{\displaystyle x^\prime} 
( p^{\displaystyle  n})_{\displaystyle x,x^\prime  
 } G_{\displaystyle 0,0} (x^\prime) \to R(x) $  for large $n$   ) 
which, as we have seen,  
is in general different from the ground state $\psi_{\displaystyle 0} (x)$
 we are interested in, due to  
 the  weights $w_n$  that weight differently the various 
configurations  $x$ distributed according to $R(x)$.  
We are allowed to  consider this 
state $R(x)$ 
as the initial trial state $|\psi_T>$ used 
 in the power method (\ref{power}), and that,  
at any
Markov   iteration $n$,  the walker had  weight  $w=1$ 
 $L$ iterations backward,  when it was at equilibrium
 according to the distribution $R(x)$, described before.
 In this way it is simple to compute the global   weight of the walker with
$L$ power method correcting factors:
 \begin{equation} \label{defgl}  G_{n}^L
=\prod\limits_{j=1}^L b_{\displaystyle x_{n-j}}  \end{equation} 
Therefore,  for instance , in order to compute the energy with  a single Markov
chain  of many iterations,    the following quantity is usually sampled:
\begin{equation} \label{defemax}
 E_{\displaystyle 0}= { \sum_n b_{\displaystyle x_n} G_n^L  \over 
\sum_n G_n^L } 
\end{equation}
with $L$ fixed.\cite{hether}
The reason to take $L$ as small as possible is that for large $L$ the weight
factors $G_n^L$ diverge  exponentially leading to uncontrolled fluctuations. 
In order to compute the variance  of the $G_n^L$ factors we can simply apply
what  we have derived in the previous section and prove in few lines the
exponential  growth of the fluctuations of the weights.
 Using Eq.(\ref{defmomk}) it is easily found  that:
$$ (\delta G^L)^2 = G_{\displaystyle 2,L} (x) 
- G_{ \displaystyle  1,L } (x)^2 $$
 According to Eq.(\ref{defitg}) $G_{\displaystyle 2,L}$ for large $L$ 
 diverges exponentially fast as $\lambda_2^{ L}$ where $\lambda_2$ 
is the maximum eigenvalue  of the matrix $p b^2$ ($b$ is here a diagonal matrix
$b=\delta_{\displaystyle  x,x^\prime} b_x $), whereas the first momentum
$G_{\displaystyle 1,L}$  
  diverges  as $G_{\displaystyle 1,L} \sim \lambda^L$, with $\lambda $  the
maximum eigenvalue of the hamiltonian matrix $H=p b$.  It is clear therefore
that we get an exponential increase of the  fluctuations 
$$(\delta G^L)^2 \sim (\lambda_2^L - \lambda^{2 L} ) $$ 
as in general $\lambda_2 >  \lambda^2 $ and the equality sign holds only if 
the matrix $b$ is a constant times the identity matrix.  

In order to overcome the problem of exponentially increasing variance, in the
following section we will discuss  a way to propagate   a set of $M$ 
walkers simultaneously. By evolving them
independently, clearly no improvement is obtained for the aforementioned large
fluctuations, as for this purpose  it is equivalent to iterate longer a single
walker. Instead, before the variance of the weights $w_i$ becomes too large,  
it is better   to  redefine  the set of walkers by dropping out the ones with
a weight which is too small, and correspondingly generate copies of the more
important ones, so that after this reconfiguration all the walkers have
approximately the same weight.
By iterating  this process 
the  weights  of  all the walkers  are kept approximately  equal  
during the simulation.
 This property yields 
 a considerable 
reduction of the statistical errors, as 
 the   variance of  the average  
weight $\bar w ={ 1\over M} \sum_i w_i$ is reduced by a factor 
$ \sqrt{M}$. 
This  allows  therefore a more stable propagation even 
for large $L$.

\section{ Carrying  many configurations simultaneously}

Given the $M$ walkers
we indicate the corresponding configurations and weights with a couple of
vectors
$( \underline{w}, \underline{x} ) $, with each vector component
$w_i,x_i\,\,\,i=1,\cdots,M$, corresponding to the $i^{\rm th}$ 
walker.
 It is then easy to generalize Eq. (\ref{procp})  to many independent walkers:
\begin{eqnarray} \label{procmany}
P_{\displaystyle n+1} ( \underline{w},\underline{x}) &=& 
\sum\limits_{\displaystyle x^\prime_1,x^\prime_2,\dots,x^\prime_M}  
P_{\displaystyle n}  ( w_1/b_{\displaystyle x_1}
 , w_2 /b_{\displaystyle x_2}, \cdots w_M/b_{\displaystyle x_M},
x^\prime_1,x^\prime_2,\cdots x^\prime_M)   \nonumber \\   
& & \left(   p_{\displaystyle x_1,x^\prime_1}
 p_{\displaystyle x_2,x^\prime_2} \cdots p_{\displaystyle x_M,x^\prime_M} 
\right) / ( b_{\displaystyle x_1} b_{\displaystyle x_2}
\cdots b_{\displaystyle x_M} ) 
\end{eqnarray}
If the evolution of $P$ is done without further restriction each walker is
uncorrelated from any other one and :
$$ P(w_1,w_2,\cdots,w_M, x_1,x_2,\cdots x_M) = P(w_1,x_1) P(w_2,x_2) \cdots 
P(w_M,x_M) $$
Similarly to the previous case we can define the momenta over the weight
variable:
\begin{equation} \label{gmomk}
G_{\displaystyle k,n} (x) = \int dw_1 \int dw_2 \cdots \int dw_M 
\sum\limits_{\displaystyle \underline{x} } \left( { w_1^k
\delta_{\displaystyle x,x_1} + w_2^k \delta_{\displaystyle x,x_2} + \cdots w^k_M \delta_{\displaystyle x,x_M} \over M }
\right) P_{\displaystyle n} (\underline{w},\underline{x}) 
\end{equation}

Since we are interested only to the first momentum of $P$ we can define a
reconfiguration process that change the probability distribution $P_n$ without
changing its first momentum, and in this we follow \cite{hether}:
\begin{eqnarray} \label{branching}
P^\prime_n ( \underline{w^\protect\prime},\underline{x^\protect\prime} ) &=& 
\int \sum\limits_{\displaystyle \underline{x}} G( \underline{w^\prime},\underline{x^\prime};
\underline{w},\underline{x} )  P( \underline{w},\underline{x}) \left[ d
\underline{w} \right]  \\
G( \underline{w^\prime},\underline{x^\prime};\underline{w},\underline{x} ) &=& 
\prod\limits_{\displaystyle i=1}^M \left( { \sum_j w_j \delta_{\displaystyle \displaystyle x^\prime_i,x_j} 
 \over \sum_j w_j  }  \right) \delta ( w^\prime_i - { \sum_j w_j \over M} ) 
\end{eqnarray}
Hereafter  the multiple integrals over all the $w_j$ variables are
conventionally  shorthand by $ \int \left[ d \underline{w} \right] $.
Note that the defined Green function $G$ is normalized
$\int \left[ d \underline{ w^\prime} \right] \sum_{\displaystyle \underline{x^\prime} } G=1$. 

In practice this reconfiguration process amounts to generate a  new set of 
$M$ walkers $(w^\prime_j, x^\prime_j)$ in terms of   the given $M$ walkers 
$(w_j, x_j)$  in the following way. Each new walker $w^\prime_j, x^\prime_j$
will have the same weight  $\bar w= { \sum_j w_j \over M}$ and an arbitrary
configuration $x^\prime_j$ among the  possible old ones $x_k$ $k=1,\cdots, M$,
chosen with a probability $ p_k= w_k/\sum_j w_j$.
It is clear that after this reconfiguration  the new $M$ walkers
have by definition the same weights 
and most of the irrelevant walkers with small
weights are   dropped out. This is just
the desired reconfiguration  which plays the same stabilization effect of
the conventional branching scheme.\cite{ceperley}
For an efficient implementation of this reconfiguration 
scheme see the Appendix.     
 
\section{  Bias control }\label{biasc}
It is well known that the control of the  population size $M$ 
introduces some bias in the simulation
simply because some kind of correlation between the walkers is introduced. 
However  for high accuracy calculations this bias 
often becomes the most difficult part to control. In this section we can instead
prove that the reconfiguration of the $M$ walkers defined in
 (\ref{branching}) does
a better job.  Though this reconfiguration clearly introduces some kind of
correlation among the walkers,  it can be rigorously proven that the first
momentum
 $G_{\displaystyle 1,n} (x)$ of the distribution 
of $P$ is exactly equal to the one $G^\prime_{\displaystyle 1,n}(x)$ 
 of $P^\prime$,  obtained after reconfiguration. 
This means that there is no loss of information in the described 
reconfiguration process  and 
 \begin{equation} \label{final} 
 G^\prime_{\displaystyle  1,n} (x) = G_{\displaystyle 1 n} (x) 
\end{equation}

\underline{Proof}: 

By definition using (\ref{gmomk} ) and (\ref{branching})  
$$G^\prime_{\displaystyle 1 n} (x)=\int \left[d \underline{w} \right] 
 \int \left[d \underline{w^\prime} \right]
\sum_{\displaystyle \underline{x},\underline{x^\prime}}  \left( { \sum_j w^\prime_j
\delta_{\displaystyle \displaystyle x,x^\prime_j} \over M} \right) 
G( \underline{w^\prime},\underline{x^\prime};\underline{w},\underline{x} )
 P(\underline{w}.\underline{x}) $$
The first term in the integrand contains a sum. It is simpler to single out 
each term of the sum 
$w^\prime_k  \delta_{\displaystyle \displaystyle x,x^\prime_k}/M$ and
to integrate  over all the possible variables
$\underline{w^\prime},\underline{x^\prime}$  but $w^\prime_k$ and $x^\prime_k$. 
It is  then easily  obtained that this contribution to $G^\prime_{\displaystyle 1
n}$ conventionally indicated as 
 $[ G^\prime_{\displaystyle 1,n} ]_k $ is given by:
$$ 
 [ G^\prime_{\displaystyle 1,n} ]_k = \int \left[d \underline{w} \right]
 \int \left[d w^\prime_k\right] 
\sum_{\displaystyle \underline{x},x^\prime_k} { w^\prime_{\displaystyle k} \over
M} \,\,\, \delta_{\displaystyle  x,x^\prime_{\displaystyle k} } 
 \left( { \sum_j  w_j  \delta_{ \displaystyle x^\prime_k, x_j} \over
\sum_j w_j } \right) \delta (w^\prime_k - { \sum_j w_j\over M} ) 
P(\underline{w},\underline{x}) $$ 
Then by integrating simply in $d w^\prime_k$ and summing over $x^\prime_k$ 
in the previous integrand we easily get that $[ G^\prime_{\displaystyle 1,n} ]_k = 
{1 \over M} G_{\displaystyle 1,n} $, independent of $k$. Finally by summing over $k$ 
we prove the statement   (\ref{final}).

\section{The GFMC scheme with bias control}
Using the previous result it is easy to generalize the equations
(\ref{defgl},\ref{defemax}), to many configurations. 
It is assumed that the reconfiguration  process described in the
 previous section
is applied iteratively each ${\bf k_b}$ steps of independent walker
propagation. 
The index $n$ appearing in the old expressions (\ref{defgl},\ref{defemax}) 
now  labels the $n^{th}$ reconfiguration  process. The measurement of the
energy can be done after   the reconfiguration  when all the walkers have the
same weight,  thus in Eq. (\ref{defgl}) :
\begin{equation} \label{bvec1}
 b_{x_n} \to b_{\underline{x}_n} \to { 1\over M} \sum_{j=1}^M b_{x_j^n}
\end{equation} 
or, for a better  statistical error, the energy can be sampled  just before the reconfiguration, taking
properly into account the weight of each single walker:
\begin{equation} \label{bvec2} 
 b_{{\underline x}_n} =
{ \sum_{j=1}^M  w_j b_{x_j^n} \over \sum_{j=1}^M w_j } 
\end{equation} 
 It is important, after each
reconfiguration,to save the quantity $\bar w={1 \over M } \sum_{j=1}^M w_j $ and 
reset to one  the  weights of each walker.  
Thus the  total  weights $G^L_n$, 
correspond to the application of $L k_b$  power method  iterations
 (as in Eq.\ref{power}) to the equilibrium distribution 
$\psi_T(x)= G_{\displaystyle  0,n-L} (x)$ , which at equilibrium is independent 
of $n$. 
 The value of the factor $G_n^L$ can be easily recovered by 
following the evolution of the
$M$  walkers in the previous $L$ reconfiguration  processes and reads:
\begin{equation} \label{defglm} 
G_{n}^L =\prod\limits_{j=0}^{L-1} \bar{w}_{\displaystyle n-j} 
\end{equation}  
where the average weight 
of the walkers $\bar w$ has been defined previously. 

An example on how the method works for the calculation of the ground state 
energy of the HM in a $4 \times 4$  lattice is shown in
 Fig. (\ref{fig1}). Remarkably our method converges very fast to the exact 
result with few correcting factors and with smaller error bars 
as compared with the original 
Hetherington's   scheme. Our method does not require the population control  
reconfiguration at 
each step, and the resulting bias is considerably reduced, in the sense that 
, without any correcting factor, our value is five times closer to the exact 
result, for the example shown in Fig.(\ref{fig1}).

\section{ Importance sampling}
One of the most important advantages of the Green function Monte Carlo
technique is the possibility to reduce the variance of the energy by exploiting 
some information of the ground state wavefunction, sometimes known a priori
on physical grounds.  In order to understand how to reduce this variance,  we just note that the  method, as described in the previous sections, is not restricted to symmetric matrices, simply because we never used this property of the hamiltonian matrices.
Following \cite{kalos} we consider not the original matrix , but 
the non symmetric one :
$$ H^\prime_{\displaystyle x^\prime,x } = \psi_G( x^\prime ) H_{\displaystyle
x^\prime, x} /\psi_G(x) $$ 
where $\psi_G$ is the so called {\bf guiding wavefunction }, that has to be as
simple as possible to be efficiently implemented in the calculation of the
matrix elements and , as we will see, as close as possible to the ground state 
 of $H$. 

In order to evaluate the maximum eigenvalue of $H^\prime$, corresponding 
obviously to the ground state of $H$, the coefficient $b_x$ is now given by:
$H^\prime$  as indicated in Eq.(\ref{defemax}), in this case $b_x$ reads:  
\begin{equation} \label{impsamp}
b_{\displaystyle x_n} = 
\sum_{\displaystyle x^\prime } \psi_G(x^\prime ) H_{\displaystyle x^\prime,
x_n} /\psi_G( x_n ) 
\end{equation}
Thus if $\psi_G$ is exactly equal to the ground state  of $H$ then, by
definition, $b_{x_n} = E_{\displaystyle 0}$, independent of $x_n$. 
This is the so called {\bf zero variance property } satisfied by the method.
Namely if the guiding wavefunction approaches an exact eigenstate of $H$ , the
method is free of statistical fluctuations.
Of course one is never in such a fortunate situation, but by improving the 
guiding wavefunction one is able to considerably decrease the error bars 
 of the energy. 
This property, rather obvious, is very important and non trivial.

For  the application of the method  to the HM
 we have used a  Jastrow like guiding function:
\begin{equation} \label{guiding} 
|\psi_G>= \sum_x s_M(x) e^{ \displaystyle {\gamma \over 2} \sum\limits_{R,R^\prime}
v(R-R^\prime )  S^z_R S^z_{R^\prime} } |x>  
\end{equation}
where $|x>$ indicate in this case all possible spin configurations with
$S^z_R=\pm {1\over2} $ defined on each site $R$ of the $l \times l$
square lattice and with the restriction of total vanishing spin  projection
$\sum_R S^z_R=0$, 
$s_M(x)$ represents the so called Marshall sign, depending on  the number 
 $N_\uparrow (x)$ 
of spin up  in one of the two sublattices
 $s_M(x)=(-1)^{\displaystyle N_\uparrow (x)} $,
while  the long range potential  is given by:
 $$v(R)= {2 \over l^2}  \sum_{q\ne 0} e^{\displaystyle i
q R} \left[1-\sqrt{ { 1+(\cos q_x + \cos q_y)/2 \over 1-(\cos q_x + \cos q_y )/2
} }  \right] .$$
with $|q_x|\le \pi $ and $|q_y|\le \pi $
 belonging to  the Brillouin zone 
 and assuming the appropriate discrete values 
of a finite system with periodic boundary conditions. 
 The constant  $\gamma$ is the only variational parameter in the wavefunction,
that  for $\gamma=1/2S$  is consistent with the spin wave theory solution 
of the HM for large spin  $S$.\cite{franjic}

\section{Forward walking}  \label{fwalking} 
The Green function Monte Carlo can be used with success to compute also 
correlation functions on the  ground state of $H$. In fact  it is 
simple to compute expectation  values of operators that are diagonal in 
the chosen basis, so that  to a given element $x$ of the basis corresponds a
well  defined value $O(x)=\langle x | O | x\rangle$ of the operator.  
By Green Function Monte Carlo, as we have seen,  configurations $w,x$ 
distributed  according to the desired wavefunction 
$\psi_{\displaystyle 0}(x)$,  or $ 
\psi_{\displaystyle 0}  (x) \psi_G (x) $ if importance sampling is 
implemented, are generated stochastically. However in order 
to compute $ <O> = <\psi_{\displaystyle 0 }  
 | O |\psi_{\displaystyle 0} > $ 
a little further work is necessary as the  
square of the wavefunction  is required to perform the quantum average.
To this purpose the desired expectation 
value is written in the following form:
\begin{equation}\label{forward}
 <O> = \lim_{N^\prime,N\to \infty}
 { <\psi_G|H^{N k_h}  O H^{N^\prime k_h} |\psi_G>  \over <\psi_G|H^{ (N^\prime 
+N ) k_h}|\psi_G> }
\end{equation}
 From the statistical point of view Eq.(\ref{forward}) amounts first to  
 sample  a configuration $x$ after $N^\prime$ GFMC reconfigurations 
, then to measure  the quantity $\langle x|O|x\rangle$ and finally to 
  let the walker  propagate forward for further  $N$ reconfigurations.

In order to evaluate the stochastic average an approach 
similar to what done for the energy is clearly possible. 
The only change   
to expression (\ref{defemax}) is to replace $b_{\displaystyle x_j}$ with  the
average measured quantity  $O_{\underline{x}_N}={1\over M} \sum_j O^n_j$ at the
generation $n$ and change the corresponding weight  factors  in (\ref{defglm})
as:  
\begin{equation} \label{defglmf}
G_{n}^L =\prod\limits_{j=-N}^{L-1} \bar{w}_{\displaystyle n-j}
\end{equation}
where henceforth we denote with $O^n_j$ the  value of the 
diagonal operator $O$ on the configuration $x_j$ of the $j^{th}$  walker, at the
iteration $n$.
  Indeed  these new factors (\ref{defglmf}) contain 
  a further  propagation of $N$ reconfiguration processes  as compared to the
previous  expression.  
It is important that both $L$,   correcting 
 the bias,  and $N$,
 correcting  the quantum average of the operator are finite, due to the 
exponential growths of the fluctuations as  $N$ and $L$ increase.
On the other hand  these fluctuations can be controlled by enlarging the
population  size $M$, and the method for $M$ large enough remains  stable. 

A further condition is however necessary in order to control the bias in
the forward walking technique.
The set of measured values $O_i^n$  with weight factors (\ref{defglmf})  has to
be modified after each reconfiguration  process occurring in the forward
direction.  In practice  after each   reconfiguration  it is important  to
bookkeep only the  values $O_i$ of the observables that 
survive after the reconfiguration (we omit in the following the superscript
$n$ for simplicity). In other words after each reconfiguration
$O^\prime_i=O_{\displaystyle j(i) }$ for $i=1,\cdots M$ 
 with the integer function $j(i)$ describing
the reconfiguration process in our scheme (after 
any reconfiguration the  walker with index 
  $i$ assumes  the configuration with index $j(i)$ before the 
reconfiguration).

In order to implement recursively the forward walking  it is 
useful to store  at each reconfiguration process the integer function $j_n (i)$
for each reconfiguration $n$ and the values $O_i$
 of the operator $O$ for each walker.
Then it is possible to compute 
  the relevant   configurations contributing to the operator 
$O$ after $N$ reconfiguration  process 
by a recursive application of the integer functions $j_n$, namely 
 $O_i^\prime=O_{\displaystyle j_N(j_{N-1} \cdots j_1(i) \cdots ))) }$.

 It is extremely important to perform the reconfiguration
process as rarely as possible since after each reconfiguration in the forward
direction the number of different configurations representing an operator $O$ 
decays quickly, yielding of course a larger variance.
 Contrary to the Hetherington\cite{hether}
algorithm our scheme allows the bias control without requiring the reconfiguration
 at each step. 

An example on how this scheme works is shown in Fig.(\ref{fig2}). 
As it is seen it is simple 
to reach the exact ground state average.

 \section{ Formal proof of the bias control in the forward walking
scheme }  \label{forwthe} 
In order to implement stochastically Eq.(\ref{forward}) we 
need to apply the operator  $O_x$ diagonal in configuration space, 
in a stochastic sense and then follow 
the standard stochastic iteration (\ref{procp})  to the walker distribution $P$
  for $N$ steps.
To this purpose a walker from now on is identified  by the triad:
$$ w, \gamma , x$$
where $\gamma$ represents the actual value of the measured
operator $O$ for the walker. Its value can change, as we will see later on 
in the reconfiguration process, and in general due to the forward walking 
$\gamma \ne \langle x |O|x \rangle$.
Indeed only at the beginning, $n=0$, of the forward walking iteration
$\gamma_i=O_i =<x_i| O|x_i>$, for $i=1,\cdots M$.
In a probabilistic sense  
this is equivalent to consider the initial  probability distribution: 
$$P_{n=0}(\underline{w},\underline{\gamma},\underline{x})  = P_0( w,x)
\prod_{i=1,M} \delta (\gamma_i -O_i) $$ where $P_0$ is the
equilibrium distribution of the previous  Markov process  (\ref{procp}), which
samples the ground state $\psi_0 (x)$. 

With this initial condition further $N$ forward walking steps are
implemented to the probability distribution $P$, defined
with the iterations  in  Eq.(\ref{procp}) and in Eq.(\ref{procmany}).
Then in order to determine the quantity (\ref{forward}) the following ratio 
is  evaluated:
\begin{equation} \label{ratioo} 
  <O> = { < w \gamma > \over < w > }
\end{equation} 
where the brackets indicate the average over the distribution $\sum_x \int dw
\int d\gamma P(w ,\gamma,x)$.
 It is understood that  in  Eq.(\ref{procp}) and Eq.(\ref{procmany}) 
the variables  $\gamma_i$ remain unchanged. 
For instance the analogous of Eq.(\ref{procp}) will be :
\begin{equation} \label{procpf}
P_{\displaystyle n+1}(w^\prime,\gamma^\prime,x^\prime)= 
\sum_x p_{\displaystyle x^\prime,x} P_n ( w^\prime/b_{\displaystyle x},
\gamma^\prime,x) /b_{\displaystyle x}
\end{equation}

 However in order to satisfy the bias control property
described in Sec.(\ref{biasc}) it is  necessary to update the
 $\gamma$ variables at any reconfiguration process. 

Analogously to the previous case it is easier to work with $w \gamma$ momenta 
of order $k$ of the distribution $P$ for fixed configuration $x$ (see
Eq. \ref{defmomk}):
\begin{equation} \label{defmomkf}
G^\gamma_{\displaystyle k,n} (x)= \int dw \int d\gamma  (w \gamma )^k
P_n(w,\gamma,x)  \end{equation}
which for $M\ne 1$ correspond to:
\begin{equation} \label{gmomkf}
G^\gamma_{\displaystyle k,n} (x) = \int d\underline{w} \int d
\underline{\gamma}    \sum\limits_{\displaystyle \underline{x} }
 \left( { \sum_j (w_j\, \gamma_j)^k  \delta_{\displaystyle x,x_j} \over  M} 
 \right) 
 P_{\displaystyle n} (\underline{w},\underline{\gamma},\underline{x}) 
\end{equation}
where, as usual underlined variables represent vectors whose components 
refer to the  single walker index $j$. 

With a proof exactly analogous to the one of Sec.(\ref{biasc})
it is possible
to show  that : 
\begin{itemize} 
\item  The   value of the first $(w \gamma)$ 
momentum $G^\gamma_{1,n} (x) $ 
, at the initial iteration of the forward walking $n=0$, is equivalent to 
apply the operator 
$O$ to the initial distribution $P_0 (w,x)$, namely
 $$G^\gamma_{\displaystyle 1,n=0}(x)  = O_x G_{\displaystyle 1,n=0} (x) $$
\item the following reconfiguration process, 
which does not change 
  the Markov chain of configurations $(\underline{w},\underline{x})$ but
 modifies slightly   $\underline{\gamma}$, 
has the bias control property also for the $w \, \gamma$ averages:
\begin{eqnarray} \label{branchingf}
P^\prime_n (
\underline{w^\protect\prime},\underline{\gamma^\protect\prime},
\underline{x^\protect\prime} ) &=&  \int \int  \sum\limits_{\displaystyle
\underline{x}} G(
\underline{w^\prime},\underline{\gamma^\prime},\underline{x^\prime};
\underline{w},\underline{\gamma},\underline{x} ) 
 P( \underline{w},\underline{\gamma},\underline{x})
\left[ d \underline{w}\right] \, \left[ d \underline{\gamma}  \right]
\nonumber\\ 
  G(\underline{w^\prime},\underline{\gamma^\prime},
\underline{x^\prime};\underline{w},\underline{\gamma},\underline{x} ) &=& 
\prod\limits_{\displaystyle i=1}^M 
\delta \left(  \gamma^\prime_i - { \sum_j w_j \gamma_j
\delta_{\displaystyle x^\prime_i,x_j } \over \sum_j w_j \delta_{\displaystyle
x^\prime_i, x_j } } \right)   
\left( { \sum_j w_j \delta_{\displaystyle
\displaystyle x^\prime_i,x_j} 
 \over \sum_j w_j  }  \right) \delta ( w^\prime_i - { \sum_j w_j \over M} ) 
\end{eqnarray}
The first factors  in the Green function $G$,  involving the $\gamma's$,
represent the only difference to the previous reconfiguration process
(\ref{branching}). Thus obviously the momenta $G_{k,n}$ not involving the
 $\gamma$ variables satisfy the same bias control property of the previous
reconfiguration process (\ref{branching}) 
 \end{itemize}

As far as the $(w\gamma)$ momenta are concerned is possible to prove as before
the mentioned bias control property: 
\begin{equation} \label{statementf}
 G^{\prime \gamma}_{1,n}(x)=G^{\gamma}_{1,n}(x)
\end{equation}

To this purpose, analogously to the previous case, it is convenient to single 
out a term $j=k$ in the definition of the first $w \gamma$ momentum in
Eq.(\ref{gmomkf}) , and following the same route of Sec.(\ref{biasc}) integrate
easily the Green function  
over all possible variables $\underline{w}^\prime$,$\underline{\gamma}^\prime$
$\underline{x}^\prime$, but the variables $x^\prime_k$, $\gamma^\prime_k$ and 
$w^\prime_k$. These remaining integrations can be also performed analytically
by first integrating in $w_k$, then in $\gamma_k$ and finally summing over
$x_k$.
The assertion (\ref{statementf})  is therefore proved rigorously.

\section{ A ``straight''~-~forward walking } \label{doublerun}
A  simpler  method to compute averages of general
operators is obtained by Eq. (\ref{ratioo}) performing 
two  independent  simulations  for  the numerator and the 
denominator in the equation (\ref{ratioo}).
The remarkable advantage of this technique is the possibility 
to measure also off-diagonal operators $O_{x^\prime,x}$. 
  After the first simulation for the evaluation of the denominator 
we take an  equilibrated walker configuration $(w, x)$ and apply the 
 operator $O$. 
Whenever the  
 operator $O$ is off diagonal this is not simply equivalent to scale 
the weight $ w \to w O_x$. In fact  we need  
a stochastic approach to select only one of the configurations
$x^\prime$ among the possible ones connected to $x$ with 
non zero matrix element  $O_{x^\prime,x}$. 
Whenever  $O_{x^\prime,x} >0$ this stochastic approach 
can be implemented  with  a  two steps technique  analogous to the one 
for the hamiltonian:   
\begin{itemize}
\item we first 
scale $w$ by the mixed average estimate  $\sum_{x^\prime} \psi_G (x^\prime) 
 O_{x^\prime,x} /\psi_G(x)  $
\item then we  select a random new configuration $x^\prime$ with a 
probability proportional to $ \psi_G (x^\prime)
 O_{x^\prime,x} /\psi_G(x)$
\end{itemize}
Then the   same reconfiguration  process (\ref{branching} ) 
  to work with a fixed number 
of walkers during the forward walking propagation can be efficiently 
applied.  

Finally we comment that the general operators  $O$ 
with arbitrary signs in the matrix elements can be always casted as 
a difference $O=O^+ - O^-$ of two operators 
with positive definite matrix elements;  the above method can be applied
to   $O^+$ and $ O^-$ separately. 

\section{Discussion and results }

In the  previous sections we have described how to obtain ground state 
energies and correlation functions of some class of   Hamiltonians 
on a finite lattice size. 
In this section we describe a successful application of this method to the 
HM. \\
We are interested in thermodynamically converged physical 
quantities characterizing the quantum antiferromagnet, for instance
the energy per site $e_0$ , the  staggered magnetization $m$ , the
spin-wave velocity $c$, the spin susceptibility $\chi$.  Use of 
a finite size scaling analysis is required to 
obtain  the infinite volume limit of our data.\\
We compute with our method 
  the ground state energy per site 
 $e_0$ and the ground state expectation value of the 
 spin-spin structure factor 
$S(q)$ :
\begin{equation} \label{czq}
S(q) = {1 \over N_a} <\psi_0| \sum_{R,R^\prime}  e^{ i q (R-R^\prime) }  S_R
 \cdot S_{R^\prime}  |\psi_0>
\end{equation}
which for  $q$ equal  to the antiferromagnetic momentum  $Q=(\pi,\pi)$ 
allows a finite size estimate
of the order parameter 
$m_l =\sqrt{  S(Q) \over N_a}$.
The  known finite size scaling theory\cite{fisher,ziman} in a quantum 
antiferromagnet predicts that 
\begin{description}
\item{a)} the ground state energy per site $e_0$ has the following leading 
size corrections:
\begin{equation} \label{efsize}
 e_0(L) = e_0    -1.4372 \; c  /l^3    + \cdots  
\end{equation} 
which allows an indirect evaluation of the spin-wave velocity.
\item{b)} 
 Further  
 the finite size estimate of the order parameter $m_l$  approaches its 
converged value as $1\over l$
\end{description}

Finally  Neuberger and Ziman \cite{ziman},  using a relativistic pion physics 
analogy,  derived very powerful constraints on the spin-spin  
structure factor $S(q)$, namely that for $q\to Q$   it diverges as 
$1/|q-Q|$ with a prefactor equal to $m^2 \over  \chi c $.
Using their arguments it also  follows immediately  that 
\begin{equation} \label{pion} 
S(q) \sim \chi c  |q|
\end{equation}
for $|q| \to 0$.

In spin wave theory results for  the constants appearing in Eqs.(\ref{efsize})  are given by:
\begin{eqnarray} \label{swconst}
m& =& s -c^\prime \\ \label{msw} 
c&=& 2 J s \sqrt{2} ( 1+c_0 /2 s ) \label{velc} \\ \label{velsw} 
\chi c &=& \frac{s}{2 \sqrt{2}} \; ( 1-c^\prime/s) \label{rhosw}   
\end{eqnarray}
where $c_0$ and $c^\prime$ can be estimated on a $l \times l$ finite size
 lattice:
\begin{equation} \label{defc0} 
c_0(l)=1- {1\over N_a} \sum_k  \epsilon_k \to 0.1579 
\end{equation} 
and 
\begin{equation} \label{defcp} 
 c^\prime(l) = \frac{1}{2N_a} \sum\limits_{k\ne 0,Q}  
\frac{1}{\epsilon_k} -\frac{1}{2} \to 0.1966 
\end{equation} 
and are expressed in terms of the spin wave energy
$\epsilon_k= \sqrt{1-\gamma_k^2}$, where   $\gamma_k=(\cos k_x +\cos k_y)/2$.

In order to improve the accuracy  of the finite size scaling calculation 
we have systematically compared our finite  size data with the spin 
wave expansion on the same lattice sizes.\cite{zhong} 
This technique allows to compute explicitly the finite size corrections 
using the  $1/S$ expansion, yielding results 
 consistent with the previous 
theory  for the finite size corrections.
 The advantage to use finite size spin wave theory is that 
 it also implicitly determines all 
 the subleading corrections in $1/l$.  
In this approach, 
the energy per site  is given by:
\begin{equation} \label{esw} 
 e(l) =-2J \left[  (s- c_0(l))^2 -1\over N_a^2  \right] 
\end{equation} 
 which correctly reproduces the predicted finite size 
scaling (\ref{efsize})  with the spin wave velocity given by Eq. (\ref{velsw}) 
and  a finite size  next leading contribution  $+2J/l^4$, appearing 
in the second order  spin wave expansion. The latter term is 
inconsistent with a claim by Fisher, which probably omitted higher 
order contributions in his analysis.\cite{fisher}  
On the other hand the fit of the data reported in Tab,\ref{taben} 
is also  in quantitative agreement   with spin wave theory, also 
 for this  next leading  contribution to the energy per site.  

By a simple Fourier transform of the 
finite size spin wave results for $< \vec S_R \cdot \vec S_0 >$ we 
obtain $S(q)=0$ for $q=0$, consistent with a singlet ground state and:
\begin{eqnarray} 
S(Q)&=& N_a (s-c^\prime(l) )^2 -1/N_a + {1\over 2 N_a} \sum\limits_{k  \ne 0,Q} 
( {\gamma_k  \over  \epsilon_k} )^2 \label{sq} \\
S(q) &=& { 1 -\gamma_q \over \epsilon_q } ( s - c^\prime (l) ) -{1 \over N_a} 
+ { 1\over 4 N_a} \sum\limits_{k \ne 0,Q,q,q+Q} { 1 -\gamma_k \gamma_{q-k}
 - \epsilon_k \epsilon_{q-k} \over \epsilon_k \epsilon_{q-k} }  
 ~~~~ q \ne Q,0  \label{sqsw2}
\end{eqnarray}
After a  simple inspection    
 the leading behavior  $S(q) \propto  |q|$  
( $S(q) \propto 1/|q-Q|$) for $q \to 0$ ($q \to Q$) is in agreement 
with the Neuberger and Ziman  predictions, with exactly consistent prefactor 
 $\chi c = { s \over 2 \sqrt{2}} ( 1 -c^\prime/s )  $ ( $m^2/\chi c = 
2 \sqrt{2} s ( 1  -c^\prime/s ) $ )  within $1\over s $ expansion.  

Now we discuss the finite size results obtained with  the GFMC
technique described in this paper. 
 
First we compute the ground state energy per site $e_0$ and 
report all   the data in 
Tab. \ref{taben}. According to Eq. (\ref{efsize}) 
we make a  
parabolic fit in $1/l$ and obtain  for the energy per site  in 
the thermodynamic limit the value reported in Tab. \ref{taben}.
This value differs  from previous GFMC estimate\cite{runge}, 
without population control  error, which is exactly removed  in 
 our method.
 This error seems to affect significantly the large size estimate of the 
energy as shown in the Tab.\ref{taben}.
 A recent paper by Sandvik\cite{sandvik},
  using a completely different path integral method,
 reports energy values perfectly consistent with ours  and with similar error 
bars.
The spin-wave velocity  is then evaluated by looking at the finite size 
corrections of $e_0$. Then according to Eq.(\ref{efsize}) we obtain:
\begin{equation}
c/c_{SW}  = 1.12 \pm 0.06    
\end{equation}  
with $ c_{SW} =\sqrt{2} J $ the zero order spin wave velocity. 
We have reached a very poor accuracy for this quantity, as it is 
determined by the subleading corrections of the energy, that in turn 
are also  quite size dependent. For this quantity  it is 
not possible to find significant  differences  from the second order 
spin wave result $c/c_{SW}= 1.1579$ (Eq.\ref{velsw}), which is probably a more 
realistic and accurate value.

The order parameter is evaluated by the forward walking technique 
with (Sec.~\ref{fwalking}) 
or without (Sec.~\ref{doublerun}) bookkeeping the branching matrix.  
In the latter case the spin isotropic square  order parameter $S(Q)/N_a$ 
is directly evaluated, as the method is not restricted to diagonal 
operators. 
As discussed in the paper the only source of systematic errors are given 
by the length of the forward walking propagation $N$ and the number 
of bias correcting factors $L$ for the ground state.  It is understood that 
for $N,L \to \infty$ our  method provides  the exact finite size value  for  
$m_l$. The finite $L$ error  is negligible, as we have used a   
large enough number of walkers  to eliminate the bias with
few correcting factors. 

The most important systematic error is due to the finite $N$. 
Its  drastic and controlled  reduction   as a function of $N$ 
  is displayed in 
 Fig.(\ref{beautiful}) which shows that 
we have achieved the converged values for $m_l$ within the error bars even 
for the largest system size $N_a=16 \times 16$.
Note also that the ``straight'' - forward method converges much more 
quickly as the total spin is conserved  in this method 
and the convergence to the ground state is determined 
by the much larger gap in the same singlet subspace.    
Our data for $m_l$  are again in perfect agreement with the 
  Sandvik's ones, who was able to obtain  much  
  smaller error bars. 
With the available data  we have extrapolated $m_l$ 
by displaying the ratio of this quantity with  
 the spin-wave prediction in Fig.(\ref{vel}).
  The suppression of the finite size effects, obtained with this analysis  
is remarkably good  (see table \ref{tabma}). 

Finally we have computed the structure factor 
 with the forward walking technique  (Sec.\ref{fwalking}) for several 
momenta. Analogously to the previous case  for the order parameter, 
 we have studied the 
ratio of the QMC data with the spin wave prediction given by Eq.(\ref{sq}).
This calculation shows that the spin wave expression (\ref{sq}) is 
particularly accurate close to $q\sim Q$ but there is some deviation for 
small momenta.   From  Eq.(\ref{sq}) the small $q$ limit of the structure 
factor can be 
computed analytically  in spin wave expansion: 
    $S(q) = |q|/ 2 \sqrt{2}  ( s -c^\prime ) = 0.1073 |q|$.
The  ratio between our QMC data and  the second order spin wave 
prediction  is  shown in Fig.(\ref{sqfig}) and for small 
$q$ it  approaches the value 
 $1.045 \pm 0.01$, rather independently from the system size. This 
 yields,
by Eq.~(\ref{pion}), a direct determination  of the product of the spin wave 
velocity and the susceptibility: 
$\chi c = 0.1122 \pm  0.001  $.
This value is very much in disagreement with Sandvik's one \cite{sandvik}  who  
predicted $ \chi c = 0.1046 \pm  0.002$, about  four error bars out from 
our direct and more accurate result. 
This discrepancy 
is probably due to a very difficult infinite size extrapolation of 
$\chi$\cite{catia} 
 and  $c$, whereas the product $ \chi \times c$ calculated
 by means of $S(q)$ 
for small momenta, seems rather well behaved, as shown in Fig.(\ref{sqfig}). 
In the same figure 
the prefactor around $q\sim Q$ is in quite reasonable 
agreement with the expected one (full big dot)  $m^2/\chi c $ obtained with 
our independent measures of $\chi c$ 
(slope $S_q$ for $q\to 0$)  and 
$m^2$ ($S(q)/N_a$ for $q=Q$), considering also that 
this function, according to the  Neuberger and Ziman's 
   theory, should jump discontinuously 
at $q=Q$, where it assumes a value determined  only by  the
 the order parameter $m^2$.
The shape of the curve displayed in Fig,(\ref{sqfig}) is clearly 
 consistent with the predicted singularity.

Our results therefore  represent a direct confirmation about the internal 
consistency of the Neuberger and Ziman's theory. 

The values  for the physical quantities extrapolated in the 
thermodynamic limit is summarized in table (\protect\ref{tabfinal}\protect).
 
\section{Conclusions}
We have described in detail a straightforward implementation of the Green 
Function Monte Carlo  scheme on a lattice hamiltonian without need of  
the standard branching process. The extension of such a scheme to 
continuum systems such as $He^4$ is straightforward.  
Indeed the present  algorithm works at fixed number of walkers and we have
shown that all sources of systematic error can be controlled with a rigorous
approach both for computing the ground state energy and for computing
ground state correlation functions. 
The possibility to work with a limited  number of walkers is extremely important
for high accuracy calculations. This in fact requires quite long simulations to
decrease the statistical errors  and, with the standard approach\cite{ceperley},
one easily exceeds the maximum number of walkers for the available 
computer memory. 

  Our reconfiguration scheme is not restricted to work at 
fixed number of walkers. In fact  our proofs in 
Sections  \ref{biasc},\ref{forwthe} can be readily generalized when   
the number $M^\prime$   
 of outgoing walkers   $x^\prime_j$ (with unit weight) 
 is different from  the number  $M$ of 
the incoming ones  $w_j$ $x_j$.
Thus a standard branching scheme between two consecutive reconfigurations 
can be also applied, and the method  can be used only  each time 
 the population of walkers reaches an exceedingly 
  large or small size. At each stochastic 
 reconfiguration  
  the simple factor $\sum\limits_{j=1}^M  w_j /M^\prime $  
corrects {\em  exactly}  the  bias of the described 
size population control. 
This maybe a more efficient implementation of our method, 
with,  however,  a rather more involved  algorithm.  

This scheme  represents a more practical implementation of the Hetherington 
idea to work at fixed number of walkers as: 
\begin{description}
\item{i)}  it is not required  to apply the 
reconfiguration process at each Markov iteration.
 Indeed our scheme coincides with the Hetherington one in this limit. 
\item{ii)}  the same idea to control the bias at fixed number of walkers 
 was extended to the ``forward walking technique'' which is important
 to calculate efficiently
correlation functions on the ground state.
\item{iii)} contrary to the conventional believe 
we have shown that there are no basic difficulties to compute 
''off diagonal'' correlation functions with Green Function Monte Carlo, 
using a simple forward walking technique (Sec\ref{doublerun}), which 
turned out to be  very  efficient  for determining 
 the order parameter in the HM. 
\end{description} 
 
We have applied these  methods 
 to compute very  accurately the ground state energy
per site of the  HM, 
the spin-spin structure factor and 
the antiferromagnetic order parameter, whose infinite size values 
are shown  in the table (\protect\ref{tabfinal}\protect). 

 We have obtained 
very good agreement with finite size spin wave theory, which allows a very 
well controlled finite size scaling (see Fig.\ref{sqfig}).
  We believe that the reported accuracy gives a very robust confirmation 
about the existence  of AF  long range order in the 2D HM.  

We finally show in Fig.(\ref{laco4}) 
a comparison of our QMC prediction for $S(q)$ 
with the  available experimental data  on the stechiometric La$_2$CuO$_4$
Mott insulator 
 for momenta close to the AF wavevector: $q \sim (\pi,\pi)$.
 The agreement is remarkably good considering also that 
\begin{description}
\item{i)} the experiments have been performed at a temperature above  the 
N\'eel temperature,  so that no true long range order exists, despite the 
very long correlation length.\cite{allen} 
As a consequence, for a good comparison between experiments and theory,
one should take into account the  smearing  
of the $\delta$ function contribution at $Q=(\pi,\pi)$,  to be added to the 
theorethical ground state prediction.
\item{ii)} there are no fitting parameters in the present comparison
\end{description}
Therefore   the Copper Oxygen planes of La$_2$CuO4, planes 
which  become   High 
Temperature superconductors upon finite  hole  doping,  are well 
described by the nearest neighbor Heisenberg model. 
\acknowledgements
It is a pleasure to acknowledge useful discussions with C. Lavalle,
L. Guidoni   and E. Tosatti. One of us (SS) wishes to thank
the Institute for Nuclear Physics of Seattle for warm hospitality.
 This work is partly supported
by PRA HTSC of the Istituto Nazionale di Fisica della Materia (SS).
\appendix
\section*{Practical reconfiguration process}

In this appendix we follow the notations  of Sec.(\ref{biasc})  to describe 
an efficient implementation of the reconfiguration process  needed 
to stabilize the proposed GFMC method.
 
 The new walkers $x^\prime_i$ after reconfiguration  are chosen 
among the old ones $x_k$, with probability $p_k$.  
We divide the interval $(0,1)$ in $M$ subintervals  with length $p_k$ 
from the leftmost  to the rightmost  as $k$ increases  from $1$ to $M$.  
Then we  generate $M$ pseudo-random numbers $z_i$ for $i=1,\cdots  M$
 and  sort them,  baring in mind  that the index $i$  labels  
 the  walker $x^\prime_i$   after the reconfiguration.
 We save therefore   the  
random permutation  $i(k)$ $k=1,\dots M$ corresponding to the described 
sorting, permutation that determines
 $z_{i(k+1)} > z_{i(k)}$. An efficient  sorting algorithm  takes 
the order of $M \log M $  operations, thus   is not time consuming.

The next step
p is to make a loop over the sorted index $k$, 
giving  monotonically 
increasing $z_{i(k)}$, 
and to select as new configuration $x^\prime_{i(k)}$  the  one $x_j$,  among 
the old configurations,   
such  that $z_{i(k)}$ belongs to the interval of length $p_j$. 
Note that the index function $j(i)$ $i=1,\cdots M$ 
 contains  all the information  required for the forward walking technique 
described in Sec.(\ref{fwalking}). 
 
After the described process  some of the   old configurations may appear 
in many copies, while others   disappear. This happens also if the 
distribution $p_j$ is uniform $p_j\sim 1/M$, yielding clearly some loss 
of information in the statistical process.
A better way to implement the reconfiguration, without loosing information 
and without introducing any source of 
 systematic error, is obtained by the following simple change,    
 After the generation 
of the random permutation $i(k)$  a new set of numbers uniformly distributed 
in the interval $(0,1)$ is defined :
$$ \bar z_i= ( \xi +(i-1) )/M$$ 
 for $i=1, \cdots M$,  where $\xi$ is another pseudorandom number in $(0,1)$.
 This set of numbers $\bar z_i$, now  uniformly distributed in the interval 
$(0,1)$,  is then used to select  the new configurations, yielding 
a more efficient implementation of the described reconfiguration 
process.  
\begin{table}
\caption{Energy per site of the HM
 in the square lattice $l\times l$ .
In this work the infinite size extrapolation is 
obtained with a parabolic 
fit $E_0(l) = E_0(\protect\infty) + a/l^3 +b /l^4$ 
for all size with $l\ge 8$. (
 $a=-2.275 \protect\pm 0.12$ and , $b=1.64\pm 0.9$).
The numbers in parenthesis represent error bars in the last digits. }
\vspace*{0.5cm}
\begin{tabular}{|l|l|l|l|} \hline \hline
$l$ & $E_0$\protect\cite{runge} &$E_0$\protect\cite{sandvik}& $E_0$ This work \\ \hline
6 &     -0.678871(8)  & -0.678873(4) & -0.6788721(28) \\ \hline
8 &   -0.673486 (14) & -0.673487(4) & -0.673483(8) \\ \hline
10 &   -0.671492(27) & - 0.671549(4) &-0.671554(6) \\ \hline
12 &    -0.670581(49) & -0.670685(5)  &-0.670678(5) \\ \hline 
14 &      =           & -0.670222(7) & -0.670223(8) \\ \hline      
16 &    -0.669872(28) & -0.669976(7) & -0.669977(8) \\ \hline
$\infty$ & -0.66934(3) & -0.669437(5) & -0.669442(26)  \\ \hline
\end{tabular}
\label{taben}
\end{table}

\begin{table}
\caption{Staggered magnetization of the
HM in the square lattice $l\times l$ with side $l$ computed
with the forward walking technique. The numbers in parenthesis represent error bars in the last digits. The extrapolated values of our GFMC
data for  $N_a \to \infty$ are obtained by the fit shown in 
Fig.\protect\ref{vel} }
\vspace*{0.5cm}
\begin{tabular}{|l|l|l|l|l|} \hline \hline
$l$ & $m_l$\protect\cite{runge} & $m_l$\protect\cite{sandvik} &
 $m_l$ This work,sec. \ref{fwalking} & This work, sec. \ref{doublerun} \\ \hline
6 & 0.4581(2) & 0.458074(3) &0.4583(3) & 0.4579(1)  \\ \hline
8 & 0.420(1) &  0.421709(9) &0.4212(6) & 0.4217(1) \\ \hline
10 & 0.397(3) & 0.399214(9) &0.3988(8) & 0.3991(6) \\ \hline 
12 & 0.378(14) & 0.38400(1) &0.380(2) &0.3834(6) \\ \hline
16 & = & 0.3647(1)  &0.361(9)& 0.3642(6)  \\ \hline
$\infty$ & 0.3075(25) & 0.3070(3) & 0.3058(12) & 0.3077(4) \\ \hline
\end{tabular}
\label{tabma}
\end{table}

\begin{table}
\caption{ Infinite size estimates of the various ground state 
quantities of the HM  discussed in the paper. The value of $m$ 
is obtained by fitting the more accurate data of \protect\cite{sandvik}
as shown in Fig.\protect\ref{vel}. The value of $\chi c$ is 
computed from the data of Fig.(\protect\ref{sqfig}),  corresponding to the 
largest size and smallest momentum.} 
\vspace*{0.5cm}
\begin{tabular}{|l|l|}
$e_0$ &  -0.669442(26) \\ \hline
$m$ &  0.3075 $\pm$ .0002 \\  \hline
$\chi c $ & 0.1122 $\pm$ 0.001  \\ \hline
\end{tabular}
\label{tabfinal}
\end{table}

\begin{figure} 
\caption{ Energy per site in a $4\times 4$ Heisenberg cluster. The continuous
line is the exact result and the  dotted (dashed) line connects GFMC data 
as a function of the number $L$ of correcting factors within our 
scheme (Hetherington's one)  described at the end of the appendix. 
 The number of walkers in this case was $M=10$ and the
reconfiguration scheme was applied each four ($k_p=4$ in the text)  iterations 
, while in the Hetherington's scheme $k_p=1$. 
The guiding wavefunction is given in Eq.(\protect\ref{guiding}) 
with $\gamma=1.2$, and all the data are obtained with the same 
amount of computer time. The insect is an expansion of the bottom-left part 
of the picture. }  
\label{fig1} \end{figure}

\begin{figure} \caption{Plot of the squared 
antiferromagnetic order parameter $m_l^2$,  
in a  $4\times 4$ Heisenberg cluster 
as a function of the number $N$ of forward walking reconfigurations. 
 The continuous  line  indicates the 
exact result. The number of  walkers was fixed to $M=20$  with 
$k_p=5$.
  The guiding function was 
the one referenced in the Fig.\protect\ref{fig1}. 
With the present technique the $z$ component of $m_l$ can be measured on 
each sampled configuration ; $m_l^z ={1 \over N_a} 
\sum_R (-1)^R S^z_R $ and spin 
isotropy is used to determine $m_l^2 = 3 < (m^z_l)^2 >$. 
} 
\label{fig2} 
\end{figure}

\begin{figure} 
\caption{ Staggered magnetization for increasing 
 lattice sizes (from top curve to bottom curve) as a function
of the forward walking iteration number N computed with the method of
section \protect\ref{fwalking} (a) and with the method of section
\protect\ref{doublerun} (b). The number of walkers for each
lattice size and figure (a) is $M = 1000$, $2000$, $3000$, $3000$, 
$3000$ with $k_{p} = 30$, $50$, $60$, $80$, $80$ and 
$l=6$, $8$, $10$, $12$, $16$ respectively, while the guiding 
wavefunction is given by Eq.(\protect\ref{guiding}) with $\gamma =
1.125$. In picture (b) the number of walkers is $M =
1000$, $2000$, $2000$, $2000$, $2000$ and $kp = 10$, $20$, $20$, $20$, $25$}
\label{beautiful}
\end{figure}

\begin{figure}
\caption{Plot of the ratio between staggered magnetization, computed by
forward walking GFMC, and the 
spin-wave staggered magnetization as a function of $1/l$. The circles  
refer to the ''straight`` forward walking technique 
 ( Sec. \protect\ref{doublerun}),  
and the squares to the data reported in Ref. \protect\cite{sandvik}. 
The lines are  weighted linear least square fit of the data. 
  The extrapolations for $l \to \infty$ are displayed in 
Tab. \protect\ref{tabma} and in Tab.\protect\ref{tabfinal} } 
\label{vel}
\end{figure}

\begin{figure} 
\caption{Plot of the ratio between the spin-spin structure factor $S(q)$, 
computed 
by forward walking GFMC,
  and the second order spin wave estimate $S_{SW}(q)$ 
 (see Eq.\protect\ref{sqsw2}). The
Structure factor has been evaluated for $L=6,8,10,12$ over the path
$\Gamma=(0,0) \to \bar M = (\pi,0) \to Y =(\pi,\pi) \to \Gamma=(0,0)$. 
 The big dot at the $Y$ point 
is the expected  $q\to Y$ limit of  $S(q) \over S_{SW}(q)$, 
by using the values  for $m^2$ and $\chi c$ reported 
in Tab.\protect\ref{tabfinal}.}
\label{sqfig}
\end{figure}

\begin{figure}
\caption{Comparison between QMC data (full squares) and experimental 
data (full dots)  for the static magnetic structure factor.
The continuous line connecting the QMC data is the second order spin wave 
prediction (see Eq. \protect\ref{sqsw2}),
 almost exact in this scale, whereas the 
line connecting the experimental data represents 
 the corresponding fit reported
in \protect\cite{allen}.} 
\label{laco4}
\end{figure}

\end{document}